\documentclass[12pt]{iopart}
\usepackage{comment}
\usepackage[english]{babel}
\usepackage[utf8]{inputenc}
\usepackage[colorinlistoftodos, color=green!40, prependcaption]{todonotes}
\usepackage{mathtools}
\usepackage{physics}
\usepackage{xcolor}
\usepackage{graphicx}
\usepackage[left=23mm,right=13mm,top=35mm,columnsep=15pt]{geometry} 
\usepackage{adjustbox}
\usepackage{placeins}
\usepackage[T1]{fontenc}
\usepackage{lipsum}
\usepackage{csquotes}
\usepackage{siunitx}
\usepackage{import}
\usepackage[pdftex, pdftitle={Article}, pdfauthor={Author}]{hyperref}
\newcommand{\cuprite}{Cu$_\mathrm{2}$O}

\begin{document}

\title[Microwave-optical spectroscopy of Rydberg excitons in the ultrastrong driving regime]{Microwave-optical spectroscopy of Rydberg excitons in the ultrastrong driving regime}

\author{Alistair Brewin$^1$, Liam A P Gallagher$^1$, Jonathan D Pritchett$^2$, Horatio Q X Wong$^1$, Robert M Potvliege$^1$, Stewart J Clark$^1$, Matthew P A Jones$^1$}
    \address{$^1$Department of Physics, Durham University, Durham, DH1 3LE, United Kingdom}
    \address{$^2$Department of Physics, King's College London, London, WC2R 2LS, United Kingdom}
    \ead{m.p.a.jones@durham.ac.uk}

\date{\today}

\begin{abstract}

We study the ultrastrong driving of Rydberg excitons in Cu$_2$O by a microwave field. The effect of the field is studied using optical absorption spectroscopy, and through the observation of sidebands on the transmitted laser light. A model based on Floquet theory is constructed to study the system beyond the rotating wave approximation. We obtain near quantitative agreement between theory and experiment across a 16-fold range of microwave field strengths spanning from the perturbative to the ultrastrong driving regime. Compared to Rydberg atoms, the large non-radiative widths of Rydberg excitons leads to new behaviour such as the emergence of an absorption continuum without ionization.

\end{abstract}

\submitto{\NJP}
\maketitle

\section{Introduction}
Strong coupling between light and matter is a significant topic in quantum physics~\cite{Frisk_2019}, with applications from quantum communication~\cite{hoi_2011, Yan_Fan_2014, Forn-Diaz_2017} and simulation~\cite{garcia_2015, Braumuller_2017} to quantum chemistry~\cite{Martinez_2018, Flick_2018}. Typically, high quality cavities are used to couple single photons to two-level quantum systems, such as atoms or artificial analogues~\cite{Forn-Diaz_2017, liu_2023}. Closely related is the topic of strong and ultrastrong driving, where large classical fields are used to drive quantum systems, resulting in non-perturbative \cite{Larson2021} optical nonlinearities such as high harmonic generation \cite{Bhattacharya2023}.  

The strong (coupling) driving regime is commonly defined to be when $\nu_0 > \Omega > \Gamma$~\cite{Flick_2018, Forn_Diaz_2019, Qin2024}, where $\hbar\nu_0$ is the bare energy gap between the excited and ground state, $\Omega$ is the (vacuum) Rabi frequency and $\Gamma$ is the relevant rate of energy loss to the environment. In this limit, nonlinear effects such as saturation and power broadening are observed \cite{Loudon2000}, while in the cavity case, a photon may be exchanged many times between the cavity and the two-level system before being lost~\cite{Kaluzny1983}. 

In contrast, ultrastrong (also called deep strong) coupling/driving (USD) occurs when the coupling to the electromagnetic field is increased further to become the largest energy scale i.e when $\Omega > \nu_0, \Gamma$~\cite{Flick_2018, Forn_Diaz_2019, Qin2024}. New effects occur in this extreme limit \cite{Frisk_2019,Forn_Diaz_2019}. In particular  the commonly used rotating-wave approximation (RWA) no longer holds, and complex dynamics take over instead of Rabi oscillations~\cite{Deng_2015}. This regime has been observed in systems where large electric dipole moments coincide with small energy level splittings, including Rydberg atoms \cite{Tommey2020}, superconducting resonators coupled to artificial atoms \cite{Niemczyk2010} and optomechanical devices \cite{Peterson2019}. As techniques for coupling quantum systems together improve, understanding the USD regime holds great promise including better quantum computation \cite{Nataf2011,Stassi2020}, control of material properties \cite{Martinez_2018,Flick_2018,Garcia2021} and microwave-optical conversion \cite{Lambert2020,Han2021}.

In Rydberg atoms, the electric dipole moment associated with transitions between neighbouring opposite-parity states scales with principal quantum number $n$ as $n^2$, while the separation scales as $n^{-3}$ providing an excellent platform for realising ultrastrong driving \cite{Meschede1985,Raimond2001,Tommey2020}. More recently, highly excited Rydberg states of excitons (bound states of an electron and hole) have also been observed in semiconductor materials \cite{Aßmann_Bayer_2020}. In cuprous oxide, \cuprite, Rydberg excitons have been observed up to $n=30$~\cite{Kazimierczuk2014,Versteegh2021}, with wave functions spanning up to $1$~$\mu$m in real space, sparking a great deal of research into the material~\cite{Heckotter2018,Ziemkiewicz2019,Orfanakis_2022}. Previous work by our group has shown that strong microwave electric dipole transitions between neighbouring states also exist for Rydberg excitons \cite{Gallagher2022}, culminating in the observation of a record microwave-optical Kerr nonlinearity \cite{pritchett2024}. In these experiments the microwave field strength was weak, enabling a quantitative description using a perturbative approach \cite{pritchett2024}.

In this paper we study the ultrastrong microwave driving of Rydberg excitons in Cu$_2$O. We measure the changes in the optical  absorption spectrum and the intensity of the microwave-induced sidebands over a range of $\sim 50$ in microwave field strength. We also develop a theoretical model to understand our results. We use non-Hermitian Floquet theory \cite{Potvliege_2024} to go beyond the RWA, and calculate the frequency-dependent susceptibility of the medium in the steady state limit. The predictions of the model are compared successfully with the experimental results up to $\Omega/\nu_0 \sim 4$, deep in the ultrastrong coupling regime (at $n=11$). The model  allows us to gain insight into workings of the system which are inaccessible to measurement. Crucially, the observed behaviour is distinct from its atomic counterpart \cite{Tommey2020} because the non-radiative broadening of the exciton states allows many excitonic states to be strongly coupled to a single microwave frequency. The agreement between experiment and theory opens exciting avenues for realising excitonic microwave-to-optical conversion with strong fields~\cite{Lambert2020,Han2021,Ziemkiewicz2023,pritchett2024}.


\begin{figure}
\centering
    \includegraphics[width=\linewidth]{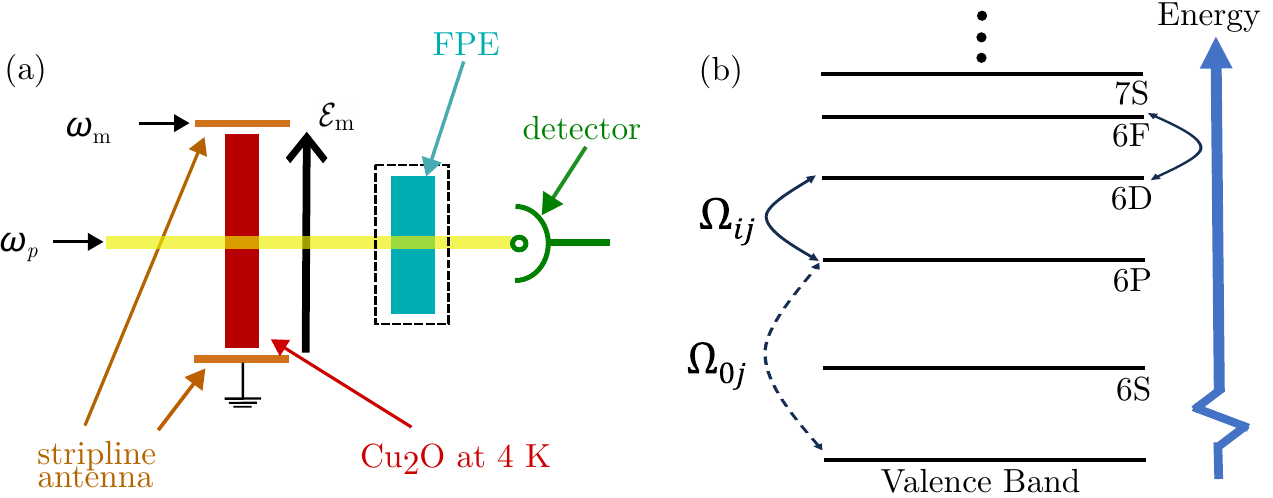}
    \caption{(a) Schematic of the experiment. A laser at optical frequency $\omega_\text{p}$ is incident on the sample, creating excitons in $n$P states. A stripline antenna driven at microwave frequency $\omega_\text{m}$ generates a strong microwave field in the sample which is cooled to $\sim$ 4~K.  A Fabry-Pérot etalon (FPE) is inserted to resolve the sidebands on the transmitted light. (b) Diagram showing the states involved in the model and their couplings. Dashed arrows denote examples of the laser field coupling the valence band to $n$P states, and solid arrows denote examples of the microwave field coupling exciton states that differ in $l$ by $1$. Here $nl$ labels the states in the same way as in hydrogen.}
    \label{fig:exp}
\end{figure}

\section{The Experiment}
\label{sec:exp}
Details of the experimental setup are available in~\cite{pritchett2024} and will be summarized here. Figure \ref{fig:exp} shows a diagram of the experiment. A thin slab of \cuprite~($55\pm10~\si{\micro\meter}$ thick) is mounted on a CaF$_2$ window inside a closed-cycle cryostat at $\sim 4 $ K.  The probe laser is a narrowband diode laser, frequency doubled to $\lambda \approx 571$~nm (details in~\cite{Rogers2022}). The laser intensity on the sample is approximately 20~\si{\micro\watt\per\milli\meter\squared} which is approximately 2 times lower than the probe intensity used in~\cite{Heckotter2021-2}, allowing interactions between excitons to be neglected~\cite{Kazimierczuk2014,Heckotter2021-2}. A microwave field is applied at a frequency of 7~GHz using a stripline antenna~\cite{Gallagher2022}. This leads to two effects: a change in the absorption spectrum, and the generation of sidebands on the laser~\cite{pritchett2024}. The probe absorption data is taken with a polarisation angle between the probe laser and the microwave field of $90^\circ$, while the sideband data is taken with the polarisation of the two fields parallel.

\begin{figure}
    \centering
    \includegraphics[width=0.49\linewidth]{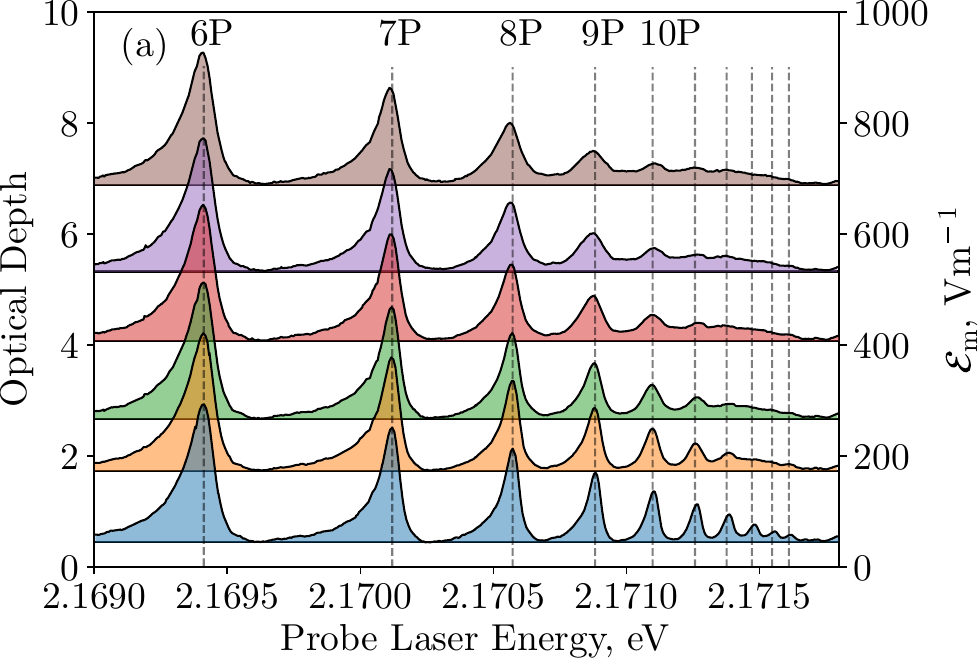}
    \hfill
    \includegraphics[width=0.49\linewidth]{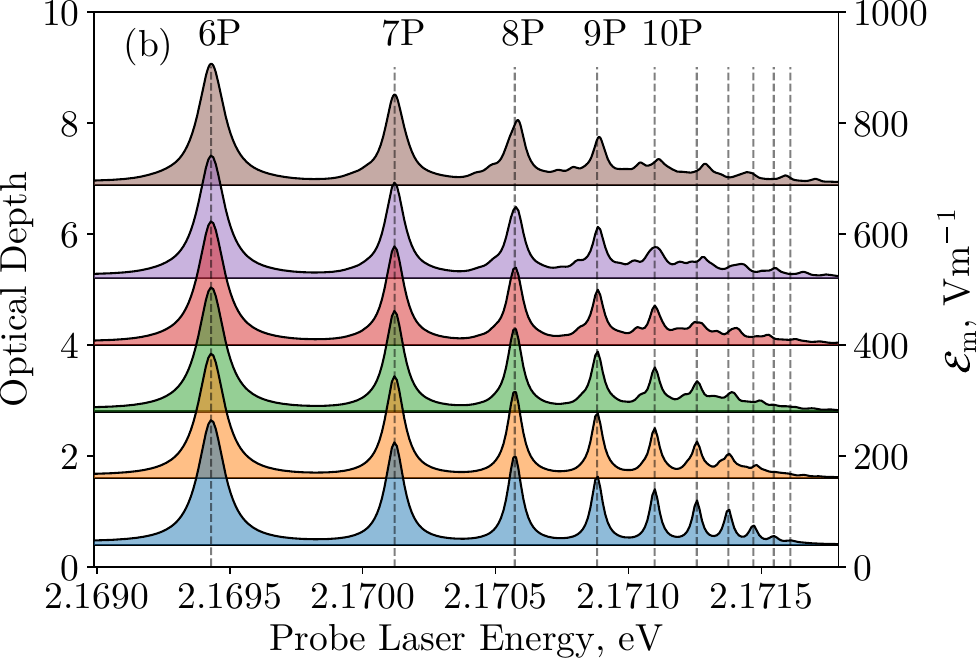}
    \hfill
    \caption{Plots of the (a) experimental and (b) theoretical absorption spectra at different microwave field strengths, $\mathcal{E}_\text{m}$. Each spectrum has been offset in optical depth by $\mathcal{E}_\text{m}/100$ to indicate the microwave field strength it was taken at, shown on the righthand y-axis. The dashed vertical lines show the positions of the zero-field $n$P resonances.}
    \label{fig:wf}
\end{figure}

Absorption spectra are taken by measuring the light transmitted through the sample, $T(\omega_\mathrm{p}$), and comparing it to a reference value, $T_0$, to give the optical depth, $-\mathrm{ln}(T(\omega_\mathrm{p})/T_0)$, as a function of probe energy (figure~\ref{fig:wf}a). Experimentally the Rydberg states are observed to sit on a non-resonant absorption background due to phonon-assisted processes~\cite{Baumeister1961}. As this background is not affected by the microwave field, it is not included in the model and is subtracted from the experimental data (see \ref{app:bgd} for details). To extract the size of the sidebands, a temperature tuned Fabry-Pérot etalon with a finesse of $44.5\pm0.7$ and a free spectral range of $60.1\pm0.2$ is inserted after the sample. Details of how the sideband amplitude is extracted from this data can be found in chapter 5 of ~\cite{PritchettThesis}.

\section{The Model}
\label{sec:model}

The calculations exploit the general non-Hermitian Floquet theory of linear absorption spectroscopy recently outlined in \cite{Potvliege_2024}. The data and code used in this paper are available at \cite{paper_data}. The Hamiltonian of our model can be thought of as the sum of a field-free Hamiltonian (figure \ref{fig:exp}b: solid horizontal lines), a coupling term between the valence band and the exciton states (figure \ref{fig:exp}b: dashed arrows), and another between pairs of exciton states (figure \ref{fig:exp}b: solid arrows). This can be written, respectively, as
\begin{eqnarray}
    \hat{H}(t) &= \,\hbar\sum_{j=1}^{\mathcal{N}}\omega^{(j)} \ket{j}\bra{j} \nonumber\\
    &-\frac{\hbar}{2}\sum_{k=1}^{\mathcal{N}}\Big[\Big(\Omega_{0k}\exp{-i\omega_\text{p} t} + \Omega_{k0}^*\exp{i\omega_\text{p} t} \Big)\ket{0}\bra{k} + \text{h.c.}\Big] \nonumber\\
    &-\frac{\hbar}{2}\sum_{j,k=1}^{\mathcal{N}}\Big[\Big(\Omega_{jk}\exp{-i\omega_\text{m} t} + \Omega_{kj}^*\exp{i\omega_\text{m} t} \Big)\ket{j}\bra{k}\Big],
\end{eqnarray}
or, making the rotating wave approximation for the probe field after transformation to slowly rotating coordinates by the unitary transformation $\hat{U} = \ket{0}\bra{0} + \sum_{j\not= 0} \exp{i\omega_\text{p}t}\ket{j}\bra{j}$,
\begin{eqnarray}
    \hat{H}_{\rm tr}(t) =& -\hbar\sum_{j=1}^{\mathcal{N}}\Big(\omega_{\rm p}-\omega^{(j)}\Big) \ket{j}\bra{j} \nonumber\\
    &-\frac{\hbar}{2}\sum_{k=1}^{\mathcal{N}}\Big[\Omega_{k0}^*\ket{0}\bra{k} + \text{h.c.}\Big] \nonumber\\
    &-\frac{\hbar}{2}\sum_{j,k=1}^{\mathcal{N}}\Big[\Big(\Omega_{jk}\exp{-i\omega_\text{m} t} + \Omega_{kj}^*\exp{i\omega_\text{m} t} \Big)\ket{j}\bra{k}\Big].
\end{eqnarray}
Here, $\ket{j}$ is an excitonic state with energy $\hbar\omega^{(j)}$
($\ket{0}$ being the ground state representing the `valence band', with $\hbar\omega^{(0)} 
= 0$) and the abbreviation `h.c.' stands for Hermitian conjugate. $\mathcal{N}$ is the number of excitonic states we include in the calculation. The Rabi frequencies between states $\ket{j}$ and $\ket{k}$ are
\begin{eqnarray}
\Omega_{0k} = \mathcal{E}_\text{p}\bra{0}\hat{D}^\text{p}_z\ket{k}/\hbar,\\
\Omega_{jk>0} = \mathcal{E}_\text{m}\bra{j}\hat{D}^\text{m}_z\ket{k}/\hbar.
\end{eqnarray} The amplitudes of the laser and microwave electric fields are given by $\mathcal{E}_\text{p}$ and $\mathcal{E}_\text{m}$ respectively, and their frequencies are $\omega_\text{p}$ and $\omega_\text{m}$, respectively. Since the linearly polarized probe laser and microwave fields define two different \textbf{z}-axes, $\hat{D}^\text{p}_z$ and $\hat{D}^\text{m}_z$ denote the $z$-component of the dipole operator in each coordinate system.

The couplings to the valence band, $\Omega_{0k}$, are derived from experiment, detailed in \ref{app:param}. The uppermost valence band of \cuprite~has even parity, so due to dipole selection rules, only odd parity states (and in particular, we assume only those with $m=0$ in the probe basis) directly couple to it. Since the F states' coupling to the valence band is several orders of magnitude weaker than that of the P states~\cite{Thewes2015}, they are difficult to measure accurately, and so are not included in this model. To evaluate the dipole matrix elements $\bra{j}\hat{D}^\text{m}_z\ket{k}$, we need wave functions for the exciton states. Rydberg excitons are commonly approximated by the Mott-Wannier model~\cite{YU_Cardona_2010}, a hydrogen-like model, with the effective masses of the electron and the hole derived from the dispersion of the conduction and valence bands as calculated with density functional theory~\cite{walther_2018}. As detailed in \ref{app:angle}, the dependence on the relative polarisation between the probe laser and microwave field predicted by this model differs significantly from experiment. Therefore to keep a self-consistent theory that is compatible with experiment, we make the following approximation:
\begin{eqnarray}
     \Omega_{jk>0} = 
     \begin{cases}
       -e\mathcal{E}_\text{m} \bra{R^{n_j}_{l_j}}r\ket{R^{n_k}_{l_k}}, &\quad\text{if } |l_j - l_k|=1,\\
       0, &\quad\text{otherwise,}
     \end{cases}
     \label{eq:dipole-no-ang}
\end{eqnarray} which gives good agreement between the model and data. To keep with convention, we still label the states $\ket{n,l}$, even though they are not  eigenstates of the $L^2$ operator. In addition, to explain the experimentally observed breaking of degeneracy of the $l$ states, we include an $l$-dependent quantum defect~\cite{schone_2016} in the exciton energies. Each different set of angular momenta $l$ then follows its own Rydberg series, $E_g - R_y/(n-\delta_l)^2$, sharing a Rydberg constant $R_y$ but offset by a quantum defect $\delta_l$~\cite{Rogers2022}. 

The excitonic states have a non-negligible energy width arising from their non-radiative decay to a state outside our basis. Taking this decay into account would normally require solving the Lindblad master equation for the relevant populations and coherences. However, since our probe is weak, we can take the population of the valence band (state $\ket{0}$) to be 1 at all times, and the populations of all the excited states to remain essentially 0, and calculate the coherences only to first order in $\mathcal{E}_\text{p}$. It can be shown \cite{Potvliege_2024} that within this weak probe approximation, the full Lindblad master equation~\cite{Marzano2020} simplifies to the following set of linear equations for the coherences $\rho_{j0}(t)$:
\begin{eqnarray}
    \fl \frac{d\rho_{j0}}{dt} = \Big(i\omega_\text{p} - i\omega^{(j)} + \frac{\Gamma^{(j)}}{2}\Big)\rho_{j0} + \frac{i}{2}\Omega_{j0} + \frac{i}{2}\sum_{k=1}^\mathcal{N}\Big(\Omega_{jk}\exp{-i\omega_\text{m}t} + \Omega^*_{kj}\exp{i\omega_\text{m}t}\Big)\rho_{k0},
    \label{eq:obe1rho}
\end{eqnarray}
where $\Gamma^{(j)}$ is the decay rate of state $\ket{j}$.
The same result would also be obtained within the weak probe approximation if these coherences were calculated by replacing the real energies $\hbar\omega^{(j)}$ by the complex energies $\hbar[\omega^{(j)}-i\Gamma^{(j)}/2]$ in the Hamiltonian $H_{\rm tr}(t)$ and solving the von~Neumann equation rather than the Lindblad master equation~\cite{Potvliege_2024}.

Accordingly, we can write the coherences $\rho_{j0}(t)$ as $c_j(t)c_0^*$ with $c_0 \equiv 1$, define
the state $\ket{\Psi(t)} = \sum_j c_j(t) \ket{j}$, and write down the equations of motion of the coefficients $c_j(t)$ using the Schr\"odinger equation for the transformed Hamiltonian. They read
\begin{eqnarray}
    \fl \frac{dc_j}{dt} = \Big(i\omega_\text{p} - i\omega^{(j)} + \frac{\Gamma^{(j)}}{2}\Big)c_j + \frac{i}{2}\Omega_{j0} + \frac{i}{2}\sum_{k=1}^\mathcal{N}\Big(\Omega_{jk}\exp{-i\omega_\text{m}t} + \Omega^*_{kj}\exp{i\omega_\text{m}t}\Big)c_k.
    \label{eq:obe1}
\end{eqnarray} We do not include an equation for the population of the valence state, $c_0$, here, since we assume that $c_0 \equiv 1$. 
Solving this system of equations shows that in the long time limit the system evolves into a steady state where the coefficients $c_j(t)$ oscillate at multiples of the microwave frequency.
Accordingly, we look for the Floquet-like solutions
\begin{eqnarray}
     c_j(t) = \sum_{N=-\infty}^{\infty} c_{j;N}\exp{-iN\omega_\text{m}t}.
    \label{eq:floq}
\end{eqnarray} where the coefficients $c_{j;N}$ are constant in time and $N$ is the Floquet number. Equating terms with equal frequency in (\ref{eq:obe1}) we find
\begin{eqnarray}
\fl \Big(\omega_\text{p} - \omega^{(j)} + i\frac{\Gamma_j}{2} + N\omega_c\Big)c_{j;N} +\frac{1}{2}\sum_{k=1}^{\mathcal{N}}\big[\Omega^{*}_{kj}c_{k;N+1} + \Omega_{jk}c_{k;N-1}\big] = -\frac{1}{2}\Omega_{j0}\delta_{N,0}, \text{ \space } j = 1,...,\mathcal{N}.
\label{eq:obe2}
\end{eqnarray} These equations can be solved numerically, with finite $N$, for the steady-state Floquet components of the wave function. All the parameters in these equations can be derived from experiment or the Mott-Wannier theory, as detailed in \ref{app:param}.

To extract the susceptibility of the medium, we need the polarisation, 
\begin{eqnarray}
     P(x,t) = n_d\bra{\Psi}\hat{D}^\text{p}_z\ket{\Psi},
    \label{eq:pol}
\end{eqnarray} where $n_d$ is the effective density of exciton states. Expanding (\ref{eq:pol}) in the manner of (\ref{eq:floq}), considering that only terms oscillating at frequency $\omega$ contribute significantly to $\chi(\omega)$, the susceptibility can be shown to take the form
\begin{eqnarray}
\label{eq:chi}
    \chi(\omega_\text{p} + N\omega_M) = \frac{n_d}{\epsilon_0 \mathcal{E}_\text{p}} \sum_{j=1}^{\mathcal{N}} \bra{0}\hat{D}^\text{p}_z\ket{j}c_{j;N}.
\end{eqnarray} For comparison with experiment, in the weak probe limit, the Beer-Lambert law, $I(x)=\exp{-\alpha x}I(x=0)$, gives the absorption of the laser at a distance $x$ into the medium for absorption coefficient $\alpha$. This is calculated from the susceptibility by
\begin{eqnarray}
    \alpha(\omega_\text{p})=2\frac{\omega_\text{p}}{c}\text{Im}\Big[\sqrt{1+\chi(\omega_\text{p})}\Big],
\end{eqnarray} where $c$ is the speed of light in vacuum. 

At the probe frequency, we study the absorption of the probe laser, but the medium also generates fields at frequencies not present in the incident light. Consequently, sidebands on the laser are generated with intensity 
\begin{eqnarray}
     I(\omega_\text{p} + N\omega_\text{m}) = \eta \big|\chi(\omega_\text{p} + N\omega_M)\big|^2,
    \label{eq:sb}
\end{eqnarray} for $N \neq 0$, and $\eta$ is a constant of proportionality which depends on the length of the crystal, the frequency of the light, the refractive index of Cu$_2$O, and the collection efficiency of the sideband light. We do not have access to the value of $\eta$ due to a lack of knowledge about collection efficiencies, and therefore we fit this parameter from the data (see section~\ref{sec:results}).

To calculate the contribution from individual Floquet states, we first recast (\ref{eq:obe2}) in terms of vectors and matrices. With $F$ as the Floquet Hamiltonian (to within a factor of $\hbar$), one can see that 
\begin{eqnarray}
     (\omega_\text{p}I - F)\boldsymbol{c} = \boldsymbol{d},
    \label{eq:obe3}
\end{eqnarray} where the components of the vectors $\boldsymbol{c}$ and $\boldsymbol{d}$ are $c_{j;N}$ and $-\Omega_{j0}\delta_{N,0}/2$, respectively. Assuming that $F$ has a complete set of eigenvectors (i.e., that $F$ is not defective),
\begin{eqnarray}
     F = \sum_q \frac{E_q}{\hbar} \boldsymbol{R}_q \boldsymbol{L}_q^\text{\textdagger}
\end{eqnarray} and
\begin{eqnarray}
     I = \sum_q \boldsymbol{R}_q \boldsymbol{L}_q^\text{\textdagger},
\end{eqnarray} where $E_q$ are the (complex) eigenenergies of $F$ and $\boldsymbol{L}_q$ and $\boldsymbol{R}_q$ are the associated left and right eigenvectors of $F$, normalised such that
$\boldsymbol{L}_q^\dagger\boldsymbol{R}_{q'}=\delta_{qq'}$. ($\boldsymbol{L}_q^\dagger\boldsymbol{R}_{q'}$ is the inner product of the column vectors $\boldsymbol{L}_q$ and $\boldsymbol{R}_q$, i.e., the matrix product of the row vector $\boldsymbol{L}_q^\dagger$ with the column vector $\boldsymbol{R}_q$, which can also be written as the dot product 
$\boldsymbol{L}_q^*\cdot\boldsymbol{R}_{q'}$; $\boldsymbol{R}_q \boldsymbol{L}_q^\text{\textdagger}$ is their outer product.) The left and right eigenvectors of $F$ need not be the identical since $F$ is not Hermitian, but they do share eigenvalues. The assumption that $F$ is not defective is in general reasonable given that this matrix is Hermitian when the widths $\Gamma^{(j)}$ are all zero. We have verified numerically that the coefficients obtained from  equation~(\ref{eq:cfinal}) below are identical to those obtained by solving equation~(\ref{eq:obe2}), which justifies this assumption in the present case. Now, substituting into (\ref{eq:obe3}) and left-multiplying by $\boldsymbol{L}_{q'}^\text{\textdagger}$ we find
\begin{eqnarray}
    \boldsymbol{L}_{q'}^\dagger \boldsymbol{c} = \frac{\boldsymbol{L}_{q'}^\dagger \boldsymbol{d}}{(\omega_\text{p} - E_{q'}/\hbar)}.
\end{eqnarray}
Finally, left-multiplying by $\boldsymbol{R}_{q'}$ and summing over $q'$, we see that the steady-state coefficients decompose into Floquet eigenstates as~\cite{Potvliege_2024}
\begin{eqnarray}
     \boldsymbol{c} = \sum_q \boldsymbol{R}_q \frac{\boldsymbol{L}_q^\dagger \boldsymbol{d}}{(\omega_\text{p} - E_q/\hbar)},
\label{eq:cfinal}
\end{eqnarray}
where we relabel $q' \rightarrow q$. Substituting this result into (\ref{eq:chi}) and reordering the sums reveals that
\begin{eqnarray}
\label{eq:superchi}
     \chi(\omega_\text{p} + N\omega_\text{m}) = \frac{-n_d}{2\epsilon_0 \hbar} \sum_q \frac{1}{(\omega_\text{p} - E_q/\hbar)}\sum_{j,k=1}^\mathcal{N} R_{j;N}^{(q)} \big[L_{k;0}^{(q)}\big]^* \bra{0}\hat{D}^\text{p}_z\ket{j}  \bra{k}\hat{D}^\text{p}_z\ket{0},
\end{eqnarray} where $L_{j;N}^{(q)}$ and $R_{j;N}^{(q)}$ are the components of the left and right eigenvectors of $F$ labelled in the $\ket{j,N}$ basis.


\section{Microwave Field Strength Calibration}

To quantitatively compare the model to experimental data we must calibrate the microwave field strength inside the medium. There is no way to independently probe the strength of the microwave field without disturbing the system. Therefore we obtain the calibration between applied microwave power and electric field strength by fitting the model to the data at low field strengths, where the model presented in section \ref{sec:model} reduces to the perturbative model used in \cite{pritchett2024}. We fit across the energy spectrum the change in absorption
\begin{eqnarray}
    \Delta\alpha L = \alpha(\mathcal{E}_\text{m})L - \alpha(\mathcal{E}_\text{m} = 0)L,
    \label{eq:delta}
\end{eqnarray} where $L$ is the thickness of the crystal. Examples of this data can be seen in figure \ref{fig:difs}. For 8 different input powers to the antenna between $0.3$ and $4.2$~mW, the model gave a calibration factor of $43.5 \pm 0.3$~V~m$^{-1}(\text{mW})^{-1/2}$. This is within the errorbar of the calibration in Pritchett et al.~\cite{pritchett2024}, but differs slightly from that work because more field strengths were used in the calibration here. It is also in reasonable agreement with finite element analysis modelling of the field in the sample~\cite{Gallagher2022}. Using the output port of the stripline antenna, we checked that the transmitted microwave power was linear with the input power over the range considered in this paper. Therefore this calibration is used for all microwave powers. The approach of calibrating at low field enables us to independently test the behaviour of  the high-field (ultrastrong driving)  regime.

\section{Results \& Discussion}
\label{sec:results}
\subsection{Probe Laser Absorption}
\begin{figure}
    \includegraphics[width=\linewidth]{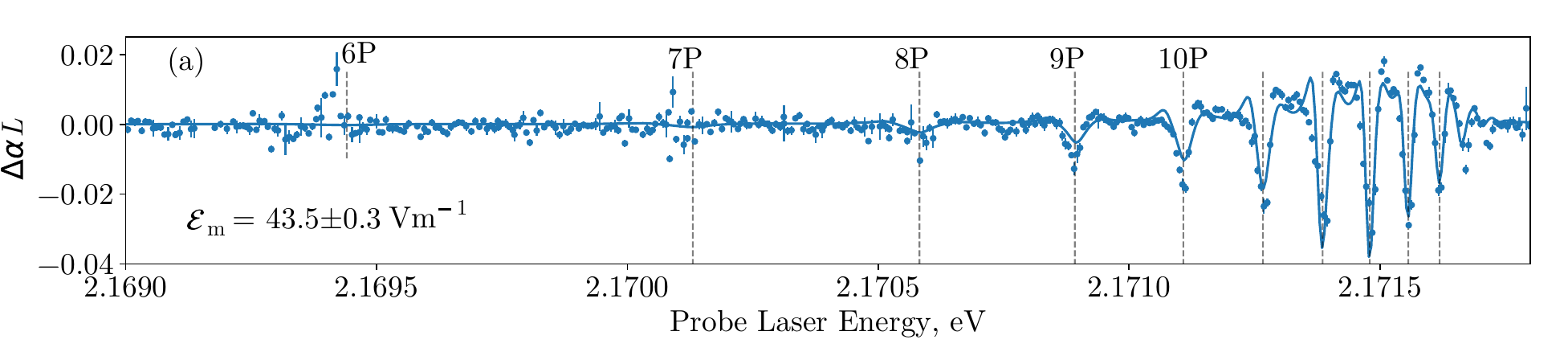}
    \\
    \includegraphics[width=\linewidth]{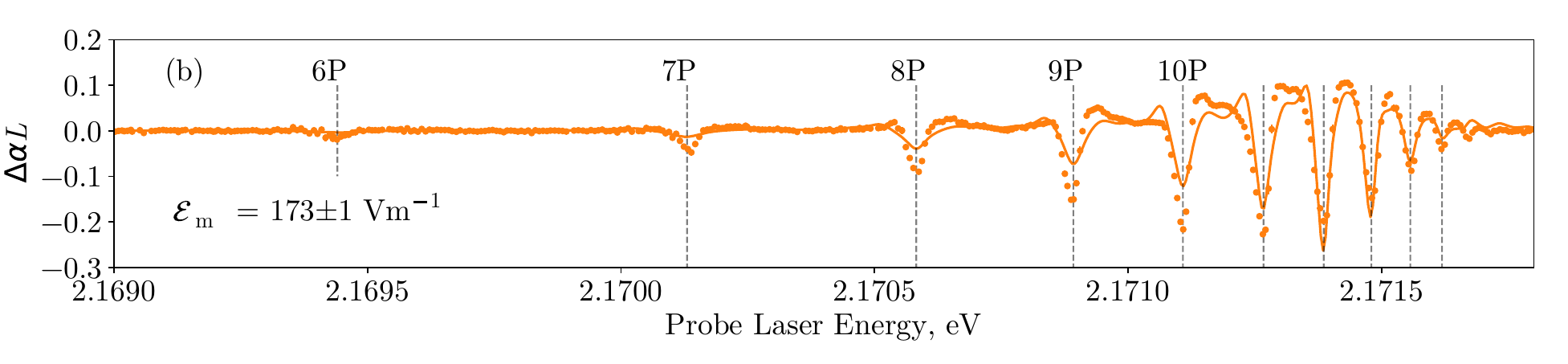}
    \\
    \includegraphics[width=\linewidth]{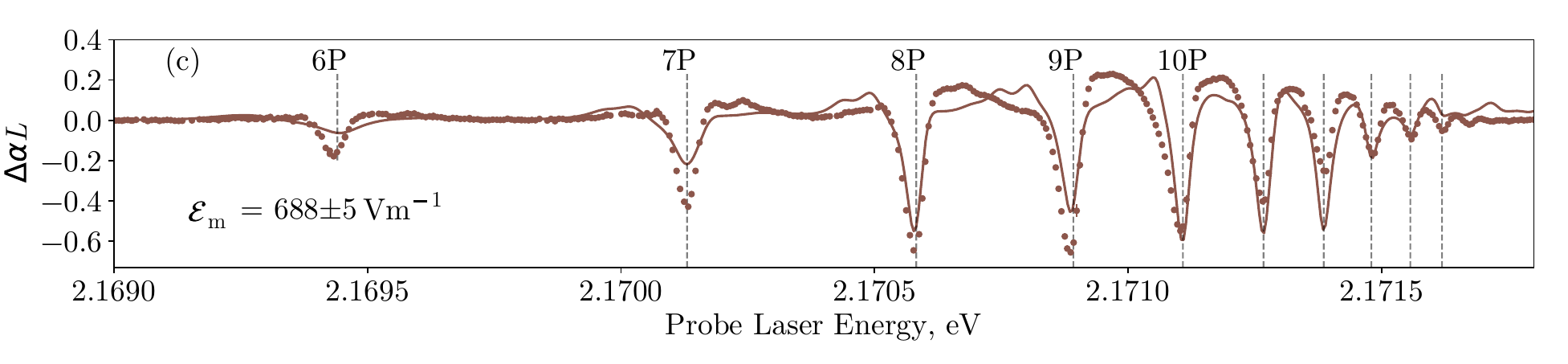}
    \caption{Comparison between the experimental (filled circles) and theoretical (solid lines) values of the change in absorption $\Delta\alpha L$ as a function of laser energy. Three different values of microwave field strength are shown spanning the perturbative (a), strong (b) and ultrastrong (c) driving regimes. The dashed vertical lines show the positions of the zero-field $n$P resonances.}
    \label{fig:difs}
\end{figure}
Figure \ref{fig:wf}a compares the experimental and theoretical results for the probe laser absorption. Each curve shows the optical depth, $\alpha L$, over the laser energy range. The curves are vertically offset in optical depth by $\mathcal{E}_\text{m}/100$ to show the microwave field strength they were taken at. As the field strength is increased, starting with the high-$n$ peaks, absorption on the peaks is reduced and absorption between the peaks is increased, resulting in a broad continuum of absorption across much of the spectrum. Qualitatively, our model reproduces these effects well. At all microwave powers, the same peaks disappear in the model as in the experimental data, with both showing a flattening of the spectrum above $9$P into an absorption continuum at the maximum field strength. In contrast to atoms~\cite{Delone_Krainov_1999}, we do not observe clear splitting and shifting of peaks in the manner of Autler-Townes in either the theoretical or experimental spectra. In experiments on atoms, Autler-Townes splitting occurs when the Rabi frequency exceeds the width of the peak, so that two peaks can be resolved. For Rydberg excitons, the non-radiative broadening of the exciton states leads to widths comparable to the state separation. Due to this, any given state is coupled to many others at once, leading to an absorption continuum.

To quantitatively compare the experimental data to the model, $\Delta\alpha L$, defined by (\ref{eq:delta}), is plotted in figure \ref{fig:difs}. Figure \ref{fig:difs}a shows the model in a region of $\mathcal{E}_\text{m}$ where it reduces to the perturbative model used in previous work~\cite{Gallagher2022, pritchett2024}. Here the agreement with data is excellent across the range of $n$, which is to be expected since this is one of the field strengths used for the microwave field calibration. Figures \ref{fig:difs}b and \ref{fig:difs}c show the theory and experiment at the threshold of ultrastrong driving and at the maximum $\mathcal{E}_\text{m}$ achieved in the experiment, respectively. Again, both show extremely good and almost quantitative  agreement a significant result for an increase in $\mathcal{E}_\text{m}$ by a factor of $16$. We stress that once the electric field is calibrated at low $\mathcal{E}_\text{m}$ (figure \ref{fig:difs}a) there are no additional free parameters. Excellent agreement is maintained through all the intermediary field strengths as well. The change in absorption is most accurately reproduced at the $n$P peaks. For reference, at $\hbar \omega_\text{p} = E_{11\text{P}}$, where $E_{11\text{P}}$ is the energy of the $11$P exciton measured from the valence band, the perturbative model predicts $\Delta \alpha L = -6$ at $\mathcal{E}_\text{m} = 688$~V~m$^{-1}$, several times the size of the absorption peak without the microwave field, indicating we are well beyond the regime where the perturbative model is valid.

 There are some discrepancies between the theory and the experiment. Note the asymmetric lineshape of the excitons in the experimental data, the precise origin of which is disputed~\cite{Toyozawa_1958, Schweiner_2016, Stolz_2018}. This was not modelled in the theory as no self-consistent method could be found. Weaker agreement between peaks can be partially explained by this lack of asymmetry in the model. However, it does not account for the spurious features in the model, for example between 8 \& 9P, or the splitting of $10$P into two peaks. The response of the $9$P resonance to the microwave field is also underestimated  by the theory. These problems are likely caused by inaccuracies in the state structure and coupling constants between excitonic states. Nevertheless, the good qualitative agreement strongly suggests the model captures the essential physics of the system.

\begin{figure}
    \centering
    \includegraphics[width=\linewidth]{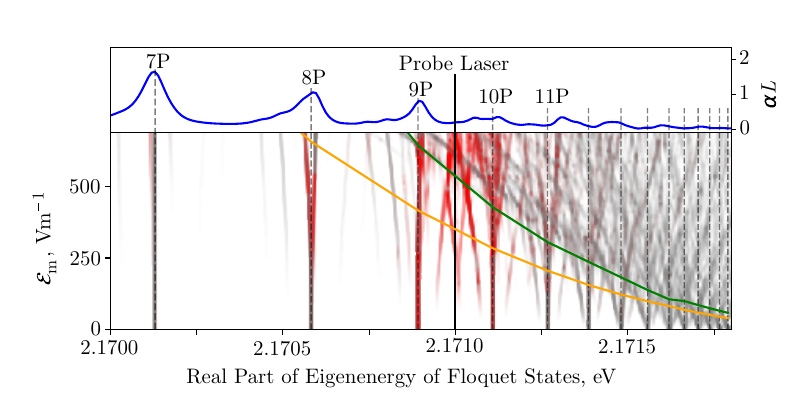}
    \caption{Bottom: a map of the real part of the eigenenergies of the Floquet Hamiltonian, Re$[E_q]$, shaded in grey proportionally to $a$ (equation (\ref{eq:a})) to demonstrate their overlap with states that couple to the valence band. The shading in red shows each state's contribution to the absorption of the laser at the indicated probe laser energy (the imaginary part of that states contribution to the susceptibility, (\ref{eq:superchi})). Above the orange curve, the $n$P to $n$D transitions enter the strong driving regime. Above the green curve, they enter the ultrastrong driving regime. Top: the theoretical absorption spectrum at $\mathcal{E}_\text{m} = 688$~V~m$^{-1}$.}
    \label{fig:stark}
\end{figure}

The model provides insight into the origin of the absorption continuum observed at high microwave field strengths. In figure \ref{fig:stark}, the real part of the eigenenergies of the Floquet Hamiltonian, Re$[E_q]$, are plotted against field strength. Grey lines indicate how strongly the state overlaps with states that can couple to the valence band, characterised by
\begin{eqnarray}
a = \sum_{n} R_{n\text{P};0}^{(q)} \big[L_{n\text{P};0}^{(q)}\big]^*.
\label{eq:a}
\end{eqnarray} As the field strength increases, more and more dressed states couple significantly to the valence band, so can contribute to the absorption. Due to the scaling of the transition dipole moments with $n$, and the widths of the eigenstates, one can see a forest of states that couple to the valence band in the high $n$ region. This forest leads to the absorption continuum observed in the experiment. As the field strength is increased, the forest can be seen to spread to lower energy, such that at the highest field strengths studied, the absorption continuum has spread as low as $9$P. Shown in red are the states that contribute to the absorption at the indicated probe laser energy, with their opacity given by the imaginary part of that states' contribution to the susceptibility, (\ref{eq:superchi}). Due to the widths of the states, at the highest field strengths shown, a broad collection of states make up the absorption at any given laser energy. The solid lines in figure \ref{fig:stark} indicate the field at which the $n$P to $n$D transitions enter strong driving (orange), where $\nu_\text{$n$P-$n$D} > \Omega_\text{$n$P-$n$D} > \Gamma_\text{$n$P-$n$D}$, and ultrastrong driving (green), where $\Omega_\text{$n$P-$n$D} > \nu_\text{$n$P-$n$D}, \Gamma_\text{$n$P}$. For example, the $11$P to $11$D transition enters the strong driving regime at $\mathcal{E}_\text{m} = 200 \pm 1$~V~m$^{-1}$ and the ultrastrong driving regime at $\mathcal{E}_\text{m} = 300 \pm 2$~V~m$^{-1}$. At the highest field strengths achieved, all states above $8$P have entered the ultrastrong driving regime, coinciding with the absorption continuum.

\begin{figure}
    \centering
    \includegraphics[width=0.6\linewidth]{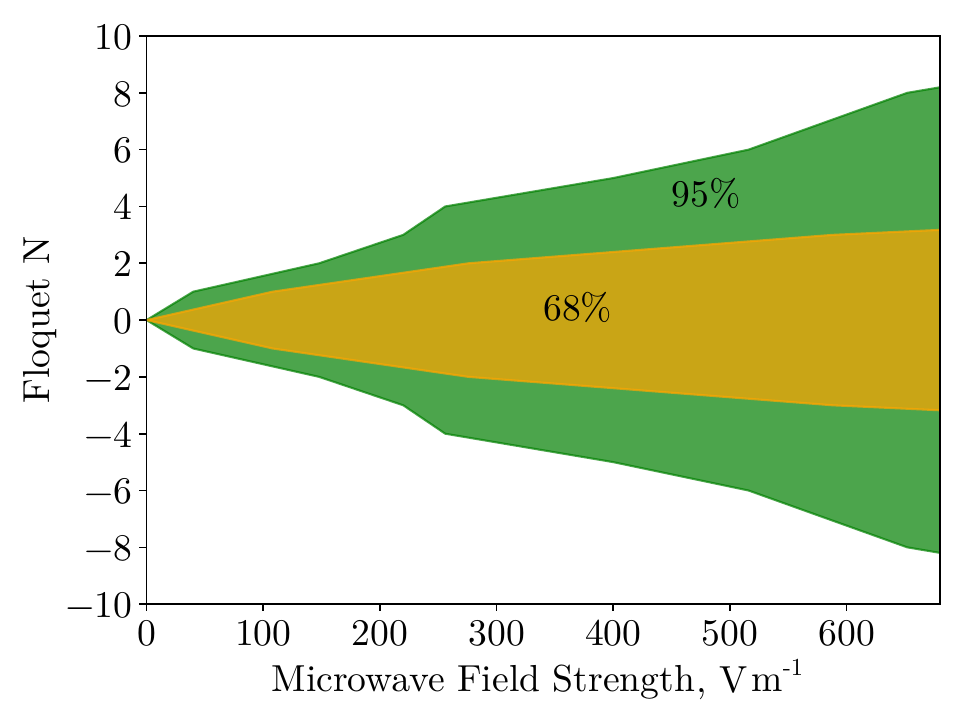}
    \caption{Contours showing the fraction of the wave function that is rotating at most at $|N|\omega_\text{m}$ plotted against microwave field strength, for probe laser energy $\hbar\omega_\text{p} = E_{12\text{P}}$. This provides a proxy for the minimum number of microwave photons involved in the system since a transition like $+\omega_\mathrm{M}-\omega_\mathrm{M}$ is a two-photon microwave transition which contributes to the $N=0$ Floquet component, but only a transition involving at least $N\omega_\text{m}$ can excite the $N$-th component.}
    \label{fig:photons}
\end{figure}

One might expect the presence of multi-microwave photon transitions in the ultrastrong driving regime. To characterize the extent of multiphoton transitions in the system, we construct the fraction
\begin{eqnarray}
     X(N) = \frac{\sum_{-N<N^{'}<N}\sum_{j\neq\ket{0}}|c_{j;N^{'}}|^2}{\sum_{N^{'}}\sum_{j\neq\ket{0}}|c_{j;N^{'}}|^2},
\end{eqnarray} which captures the proportion of the wave function rotating at a maximum frequency of $|N|\omega_\text{m}$. How $X(N)$ varies with microwave field strength is displayed in figure \ref{fig:photons}. We see clearly that the Floquet approach is justified, as higher than second order components become significant above field strengths of $200 \pm 1$~V~m$^{-1}$, around where we enter the ultrastrong driving regime. One can also see there is a significant fraction of the wave function oscillating at $|N| > 5$ at the high end of considered $\mathcal{E}_\text{m}$, well into the ultrastrong driving regime.

\subsection{Sideband Generation}

\begin{figure*}
    \includegraphics[width=0.49\linewidth]{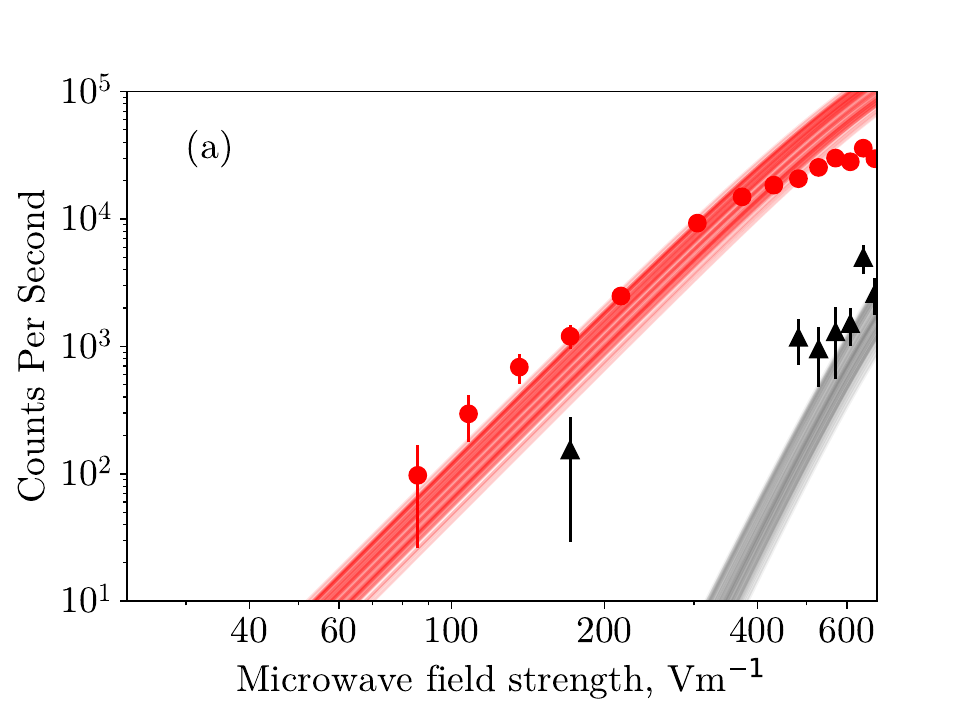}
    \hfill
    \includegraphics[width=0.49\linewidth]{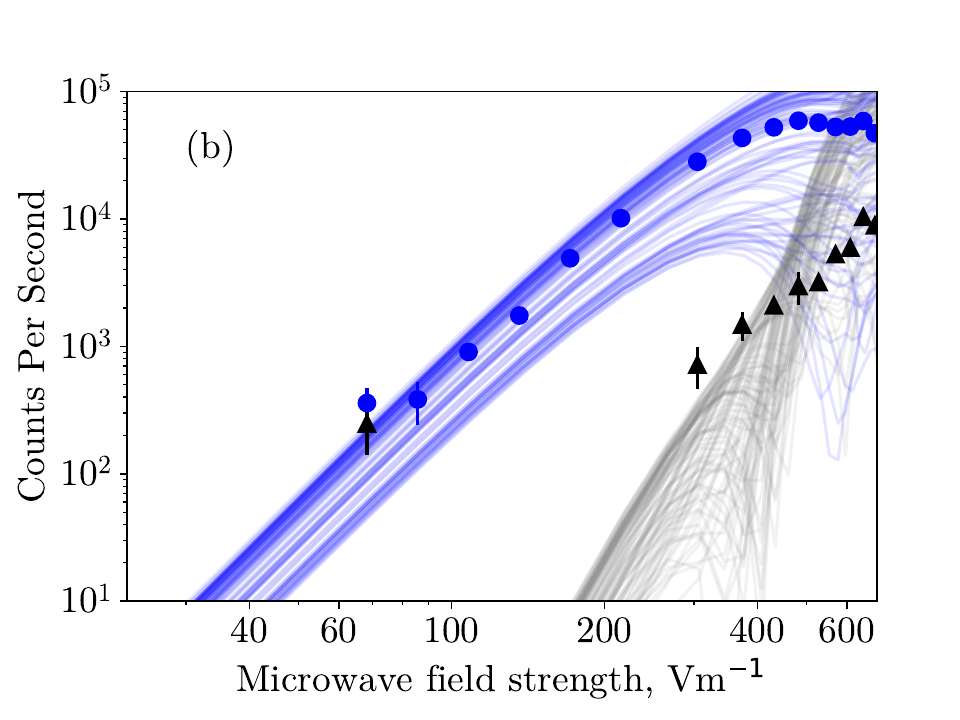}
    \hfill
    \includegraphics[width=0.49\linewidth]{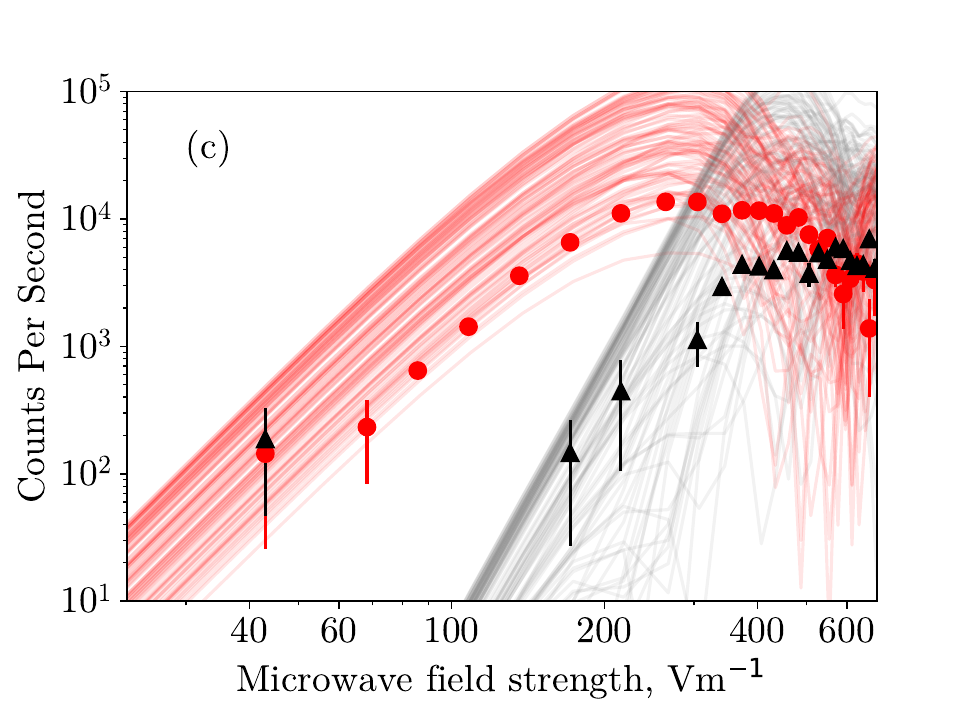}
    \hfill
    \includegraphics[width=0.49\linewidth]{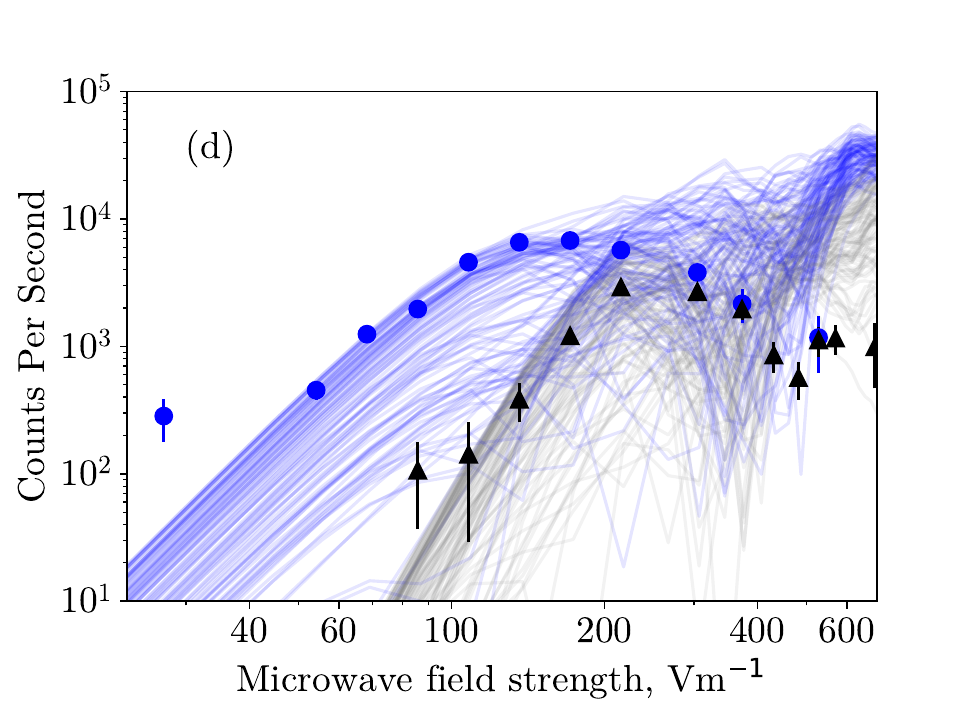}
    \hfill
    \caption{The intensity, expressed in counts per second, of second (colour) and fourth (black) order red- and blue-detuned sidebands for theory (lines) and experiment (points) at four different probe laser detunings. They are taken at probe laser energies of (a) $2.170606$~eV, (b) $2.170912$~eV, (c) $2.171252$~eV, (d) $2.171366$~eV. Each theory curve is obtained with a different, randomly-chosen, set of phases for the valence band to $n$P dipole matrix elements, $\bra{0}\hat{D}_z\ket{j}$, since these phases could not be determined from experiment.}
    \label{fig:sb}
\end{figure*}


In figure \ref{fig:sb}, the sidebands predicted by the model are compared to experiment. As mentioned in \ref{app:param}, we do not have access to the relative phases of different dipole elements $\bra{0}\hat{D}_z^\text{p}\ket{j}$, so to investigate their relevance, $100$ sets of uniformly distributed random phases were generated and the resulting carrier and sideband information plotted. Largely independent of the choices of phase, the model accurately reproduces the shape and onset of the rollover of both the second- and fourth-order sidebands over almost two orders of magnitude in field strength and nearly three orders of magnitude in sideband amplitude. For any given energy, there are even some choices that quantitatively reproduce the sidebands, providing good evidence that the model is capturing the essential physics of these processes. The quantitative agreement is better at lower laser energies, but weakens as the dependence on phase increases for higher energies. This is to be expected, since the phase dependence is introduced through cross-terms between different pairs of matrix elements in the susceptibility. Interference between the contributions of different excitons is stronger at higher $n$ where states are closer together in energy. There is no consistent choice of phase that works for all energies, suggesting an energy-dependent phase, possibly coming from the phonon background maybe be present. Taking the phase of these dipole moments into account would introduce a excessively large set of fitting parameters, detracting from the simplicity of the model. The choice of phases made little meaningful impact on the carrier absorption spectrum, so were taken to all be positive real for that analysis.

To relate the sideband intensity predicted by the model to the experimental data, the scaling parameter $\eta$, appearing in (\ref{eq:sb}), is fitted, due to the lack of access to photon collection efficiency in the experiment. The scaling parameter was fitted from the $\hbar\omega_\text{p} \approx \text{E}_{8\text{P}}$ 2nd order sidebands (figure \ref{fig:sb}a), since the $8$P exciton is relatively far away from neighbouring states. Here, the sum in (\ref{eq:chi}) is dominated by one term, i.e. that sidebands are generated principally by decay from just the $8$P state, removing dependence on cross terms between different excitons, and so dependence on their relative phases.

\subsection{Discussion}

The ultrastrong driving regime is defined as when the Rabi frequency exceeds the width of and energy spacing between the two considered levels. Although we cannot treat this system as a two-level one, it is still useful to calculate the ultrastrong driving regime condition. For example, we find that for the $11$P to $11$D transition we enter the ultrastrong driving regime for field strengths above $300 \pm 2$~V~m$^{-1}$. The point where all $n$P to $n$D transitions reach the strong and ultrastrong driving regimes can be seen in figure \ref{fig:stark}. At the highest fields applied, we can reach an ultrastrong driving parameter of up to $U = \Omega_\text{$13$P,$13$D}/\nu_\text{$13$P-$13$D} \sim 4$ for the $13$P to $13$D transition, significantly exceeding parameters of $U = 1$~\cite{Tommey2020} and $U = 2.1$~\cite{Deng_2015} in other works. It is worth noting that the strong driving regime, where the Rabi frequency is larger than the state linewidth, is reached at a similar microwave field strength, as the state linewidth (FWHM of $40~\si{\micro\electronvolt}$ for 11P) is comparable to the energy separation ($E_{11\text{D}}-E_{11\text{P}}=58~\si{\micro\electronvolt}$). 

The system presented here is qualitatively different to the analogous case of atoms. We see rapid hybridization and the creation of a quasi-resonant continuum of Floquet states, rather than Autler-Townes splitting and shifts. This continuum maintains a strong coupling to the microwave field, as shown by the measurements of the microwave-induced sidebands. The near-quantitative agreement means that this model can be used to predict the behaviour of devices used for microwave-optical conversion well beyond the linear regime and deep into the ultrastrong driving regime, maximising the exploitation of the Rydberg mediated microwave-optical nonlinearity.

One thing which is not considered by the model presented here is the effect of exciton-exciton interactions. Previous work on interactions between Rydberg states has studied the van der Waals interactions~\cite{walther_2018} which scale with exciton separation $R$ as $R^{-6}$. These experiments looked for exciton density dependent effects by varying the intensity of the optical excitation. Here, our intensity is $20$ \si{\micro\watt}mm$^{-2}$ which is $50$ time smaller than the intensity at which interactions have been seen at $n=12$~\cite{Kazimierczuk2014} and so van der Waals interactions can be neglected. However, when the microwave field is applied, the excitons states become hybridized and resonant dipole-dipole interactions can occur. These interactions are longer range than van der Waals, scaling as $R^{-3}$ and so they may be present at lower exciton densities. Some of the disagreement between the experiment and the model at high field strength may be due to these interactions.

At the largest field strengths applied, we  must consider the relevance of microwave field induced ionization, though it isn't included in the model. This is analogous to ionization in atoms, since the electron is being promoted to a continuum state, which in the case of excitons is in the conduction band. In static fields, the ionization field threshold has been observed to follow an $n^{-4}$ trend~\cite{Heckotter2018-1,Heckotter2018-2,HeckotterThesis}. These results predict that the static field required for ionization would be $\sim6000$~V~m$^{-1}$ at $n=11$, which is almost an order of magnitude larger than the highest field strength used in this work. However, the case of microwave field induced ionization is more complicated than the static field case as multi-photon absorption processes can also lead to ionization~\cite{Gallagher1994ionization}. In atomic Rydberg systems, hydrogen-like atoms follow an $n^{-5}$ scaling when the microwave frequency is less than the level spacing, and ionize at lower microwave field strengths than corresponding static fields~\cite{gallagher1994}.

Our model provides insight into whether multi-photon ionization is significant. As previously discussed, the Floquet components indicate the minimum number of microwave photons involved at a given field strength (figure \ref{fig:photons}). This information can be compared to the binding energy of the excitons. At $\hbar\omega_\text{p}=E_{12\text{P}}$ the highest Floquet component within the $95\%$ contour is $N=10$ which leads to a frequency of $70$~GHz compared with the binding energy for $12$P of $117$~GHz implying that multi-photon ionization is not significant. However, by $\hbar\omega_\text{p}=E_{15\text{P}}$ the binding energy of $15$P has dropped to 93~GHz and the maximum significant Floquet component has increased to $N=17$ within $95\%$, a frequency of $119$~GHz, so microwave field induced ionization starts to become important. A major difference from atomic Rydberg systems is the lifetime of the states. The excitonic lifetime is less than $1$~ns due to phonon scattering, which is comparable to the oscillation period of the microwave field, whereas in atomic Rydberg systems the state lifetime is much larger. Here, excitons could spontaneously decay before absorbing enough photons for ionization. Extending the theoretical model to include a continuum would allow microwave field induced ionization to be studied further, however, the addition of ionization via an arbitrary number of microwave photons is non-trivial~\cite{shakeshaft_86}. For the field strengths considered here, microwave field ionization is not expected to be playing a significant role.

\section{Conclusions}

We have constructed a two-field model describing a probe laser field and a microwave field acting on an ensemble of non-interacting Rydberg excitons in cuprous oxide, and calculated its steady state using Floquet theory, a method not previously applied systems of the kind considered here. The model quantitavely reproduces the changes in probe laser absorption across a factor of $16$ increase in microwave field strength. We are confident that, rather than ionize the excitons, the microwave field hybridizes many exciton states together, creating a continuum of dressed states. This does not occur in atomic Rydberg systems due to the non-radiative broadening of the exciton lines. The excitons are ultrastrongly driven by the microwave field, with Rabi frequencies greater than four times the energy spacing of the respective levels, where processes involving in excess of 6 microwave photons are significant. Sideband generation is well predicted and understood near low energy peaks where states are far apart, but the prediction is poorer where many states are involved, owing at least partly to unknown phases of dipole matrix elements.
\\

\section{Code and Data Availability}
The data and code used in this paper are available at \cite{paper_data}.


\ack
The authors thank V. Walther (Purdue University) for calculating dipole matrix elements for transitions between Rydberg states and I. Chaplin and S. Edwards (Durham University, Department of Earth Sciences) for preparing samples. This work was supported by the Engineering and Physical Sciences Research Council (EPSRC), United Kingdom, through research Grant Nos. EP/P012000/1, EP/X038556/1, EP/T518001/1, EP/W524426/1. 


\appendix

\section{Removal of absorption background}
\label{app:bgd}
In the absorption experiment, the absorption due to excitons sits on top of a large background of absorption due to the laser exciting a low lying exciton state and an optical phonon. This background was removed from the data as modelling its microscopic origin is beyond the scope of this work. The background absorption coefficient $\alpha_\text{bg}(E)$ is based on the function used in~\cite{HeckotterThesis}, consisting of a piece-wise function containing a below- and an above- band gap contribution:

\begin{eqnarray}
 \alpha_\mathrm{bg}(E)=\begin{dcases}
       \alpha_\mathrm{Ph}(E) + \alpha_\mathrm{U}(E) &E<E_\mathrm{g} \\
        \alpha_\mathrm{cont}  &E\ge E_\mathrm{g} \\
    \end{dcases}.
\end{eqnarray}
Here, $\alpha_\mathrm{Ph}(E)$ is the contribution to the background absorption coefficient due to phonon-assisted processes given by
\begin{eqnarray}
 \alpha_\mathrm{Ph}(E) = \alpha_\mathrm{1Sy}^{\Gamma_3^-}(E) +\alpha_\mathrm{1Sy}^{\Gamma_4^-}(E) +\alpha_\mathrm{2Sy}^{\Gamma_3^-}(E)+ \alpha_\mathrm{1Sg}^{\Gamma_3^-}(E),
\end{eqnarray}
with each $\alpha_x^y(E)$ corresponding to the contribution from the phonon assisted absorption due to the $x$ exciton state and $y$ phonon, given by
\begin{eqnarray}
    \alpha_x^y(E) = \begin{dcases}
    A_x^y \sqrt{E-(E_x+E_y)} & E> E_x + E_y\\
     0 & E\le E_x + E_y\\
    \end{dcases},
\end{eqnarray}
where $A_x^y$ is the amplitude of the contribution to the phonon assisted absorption.

\begin{figure}
    \centering
    \includegraphics[]{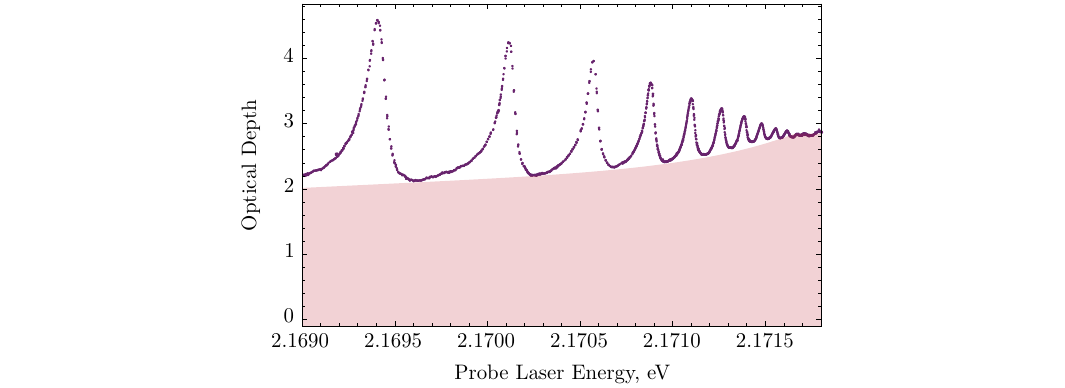}
    \caption{Removal of absorption background from experimental data. Shaded region indicated fitted background which subtracted from the experimental data in figure~\ref{fig:wf}.}
    \label{fig:bkg}
\end{figure}

The Urbach tail, $\alpha_\mathrm{U}$ is a phenomenological term which describes the exponential increase in absorption as the continuum is approached, given by
\begin{eqnarray}
     \alpha_\mathrm{U}(E)=c_\mathrm{U} \, \text{exp}\{(E-E_\mathrm{g})/E_\mathrm{U}\},
\end{eqnarray}
with amplitude $c_\mathrm{U}$ and Urbach energy $E_\mathrm{U}$. As the experimental data in this paper does not extend far above the band gap, the absorption of the continuum is treated as constant. The energies and relative amplitudes of the phonon absorption were taken from~\cite{HeckotterThesis} while the parameters in the Urbach tail were used as fitting parameters (as it is known the Urbach tail depends on temperature and the presence of defects in the material~\cite{Kruger2020}). Figure~\ref{fig:bkg} shows the absorption spectrum with the shaded region indicating the background. The parameters used for this background function are available in~\cite{paper_data}.

\section{Parameters for the model}
\label{app:param}
The energies, widths, and peak heights of the $\ket{n\text{P}}$ states are obtained for our sample from single-photon excitation~\cite{Lynch2021}. Since the $\mathcal{E}_\text{m} = 0$ carrier spectrum reduces to a sum of Lorentzians, the heights of these peaks were used to derive the magnitudes of $\bra{0}\hat{D}_z\ket{j}$. However, it affords us no information about their relative phases, so for the absorption spectrum they were all taken to be positive real. The energies and widths of the S, D and F states come from two-photon excitation~\cite{Rogers2022}, which is much less reliable due to additional broadening in the SHG spectrum~\cite{GallagherThesis}. The energies were extended beyond what was resolvable in the absorption spectrum (i.e. from $n=\{6,\ ...,\ 15\}$ to $n=\{6,\ ...,\ 20\}$) by fitting. We included the Floquet states $N = {-15,...,15}$ in the calculation to achieve a convergence of less than $0.01$ in optical depth across the spectrum.

The dipole matrix elements derived from experimental peak heights enter into the model as the product
\begin{eqnarray}
     n_d |\bra{0}\hat{D}^\text{p}_z\ket{n\text{P}}|^2 = \frac{2A_{n\text{P}}\epsilon_0}{kL},
\end{eqnarray} where $A_{n\text{P}}$ are the experimental peak heights in optical depth, $k$ is the wavenumber of the laser at that peak, and $L$ is the thickness of the sample. This makes the results of the model independent of $n_d$. It is also obvious here that we do not have access to the phases of $\bra{0}\hat{D}^\text{p}_z\ket{n\text{P}}$, only their magnitude, which is very important for the calculation of the sidebands, but less so for the absorption spectra, where they are all taken to be positive real.

\section{Angle Dependence}
\label{app:angle}

\begin{figure}
    \centering
    \includegraphics[width=0.6\linewidth]{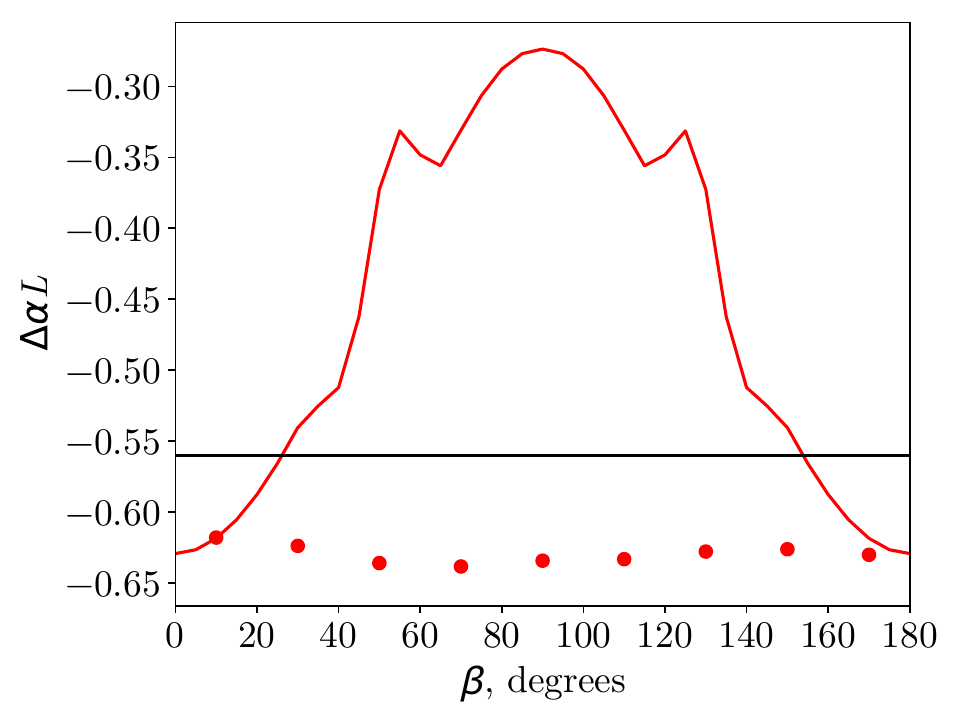}
    \caption{A plot of $\Delta \alpha L$ against the relative polarisation angle between the probe laser and microwave field, $\beta$, at $\hbar \omega_\text{p} = E_{11\text{P}} = 2.171267$~eV and $\mathcal{E}_\text{m}=688 \pm 5$~V~m$^{-1}$. The dots show the angle dependence observed in experiment, the red line shows the angle dependence predicted by the model if we take the intra-excitonic dipole matrix elements, $\Omega_{jk>0}$, from (\ref{eq:dipole-ang}), and the black line shows the model given $\Omega_{jk>0}$ from (\ref{eq:dipole-no-ang}).}
    \label{fig:angles}
\end{figure}

Mott-Wannier excitons can be separated into radial and angular parts, $\ket{\Psi_{nlm}}_\text{p} = \ket{R^n_l}\ket{Y^l_m}_\text{p}$, where the angular part is given by spherical harmonics, and the subscript p means the $m$ label refers to the probe system of coordinates. Expressing the states in the probe system as a sum of states in the microwave system,
\begin{eqnarray}
     \ket{R^{n_j}_{l_j}}\ket{Y^{l_j}_{m_j}}_\text{p} = \ket{R^{n_j}_{l_j}} \sum_{m'}d^{l_j}_{m_j,m'}(\beta)\ket{Y^{l_j}_{m'}}_\text{m},
\end{eqnarray}  where $d^{l}_{m,m'}(\beta)$ the Wigner small-d rotation matrices~\cite{Wigner1959}, $\beta$ is the angle between the probe laser and microwave field polarizations, and the subscript $\text{m}$ means the $m$ label refers to the microwave system of coordinates. Thus
\begin{eqnarray}
     \Omega_{jk>0} = -e\mathcal{E}_\text{m} \bra{R^{n_j}_{l_j}}r\ket{R^{n_k}_{l_k}}
    \times \sum_{m'm''}[d^{l_j}_{m_j,m'}(\beta)]^*d^{l_k}_{m_k,m''}(\beta) \bra{Y^{l_j}_{m'}}_\text{m} \cos{\theta_\text{m}} \ket{Y^{l_k}_{m''}}_\text{m}
    \label{eq:dipole-ang}
\end{eqnarray} where $\{r,\theta_\text{m},\phi_\text{m}\}$ are the spherical coordinates in the microwave basis to be integrated over and $e$ is the charge of the electron. However, as shown in figure \ref{fig:angles}, (\ref{eq:dipole-ang}) predicts a strong variation of the probe absorption with $\beta$ that is not observed experimentally. The full cubic symmetry of the crystal lattice does allow for the weaker dependence observed in figure \ref{fig:angles}. However the exciton wave function is no longer separable, and the Mott-Wannier picture no longer applies, hence the approximation made in (\ref{eq:dipole-no-ang}).

In figure \ref{fig:angles}, we plot $\Delta \alpha L$ against polarisation angle for $\Omega_{jk}$ given by (\ref{eq:dipole-ang}). It can be seen that, while the polarisation angle has little impact on the experimental data, (\ref{eq:dipole-ang}) predicts a very strong dependence. Moreover, the two-photon optically active S and D states have $\Gamma_5^+$ symmetry~\cite{Heckotter2021}, implying the angle dependence will be the same for both states, whereas in the atomic model S and D state have very different angle dependencies. Assigning an angular dependence of the dipole matrix elements based on a representation of the excitons wave functions by way of eigenfunctions of the angular momentum operator $\hat{\bf L}^2$ is therefore unwarranted.

\newpage
\section*{References}
\bibliographystyle{iopart-num.bst}
\bibliography{main}
\end{document}